\documentclass[aps,pra,twocolumn,superscriptaddress,showpacs]{revtex4}
\usepackage{graphicx}
\usepackage{amsmath,amssymb,color}

\def\vct#1{{\mathchoice{\mbox{\boldmath$#1$}}{\mbox{\boldmath$#1$}}%
  {\mbox{\scriptsize\boldmath$#1$}}{\mbox{\scriptsize\boldmath$#1$}}}}
\def\degree{\mbox{$^\circ$}}
\def\U#1{{%
\def\O{\mbox{O}}
\def\u{\mbox{u}}
\mathcode`\u=\mu
\mathcode`\O=\Omega
\mathrm{#1}}}
\def\ii{{\mathrm{i}}}
\def\ee{{\mathrm{e}}}

\def\bra#1{\langle #1|}
\def\ket#1{|\mbox{$#1$}\rangle}

\def\bracket#1{\langle\mbox{$#1$}\rangle}
\def\bracketi#1#2{\langle\mbox{$#1$}|\mbox{$#2$}\rangle}
\def\bracketii#1#2#3{\langle\mbox{$#1$}|\mbox{$#2$}|\mbox{$#3$}\rangle}

\def\sub#1{_{\scriptsize\mbox{#1}}}
\def\Re{\mathop{\mathrm{Re}}}
\def\Im{\mathop{\mathrm{Im}}}

\begin{document}

\title{Stereographical visualization of a polarization state using weak
measurements with an optical-vortex beam}

\author{Hirokazu Kobayashi}
\email{kobayashi.hirokazu@kochi-tech.ac.jp}
\affiliation{Department of Electronic and Photonic System Engineering,
Kochi University of Technology, Tosayamada-cho, Kochi, Japan.}
\author{Koji Nonaka}
\affiliation{Department of Electronic and Photonic System Engineering,
Kochi University of Technology, Tosayamada-cho, Kochi, Japan.}
\author{Yutaka Shikano}
 \email{yshikano@ims.ac.jp}
 \affiliation{Research Center of Integrative Molecular Systems (CIMoS), Institute for Molecular Science, Okazaki, Aichi, Japan.}
 \affiliation{Institute for Quantum Studies, Chapman University, Orange,
 California 92866, USA.}
\date{\today}

\begin{abstract}
We propose a stereographical-visualization scheme for a polarization state by
two-dimensional imaging of a weak value with a single setup.
The key idea is to employ Laguerre--Gaussian modes or
an optical vortex beam for a probe state in weak measurement.
Our scheme has the advantage that we can extract information on the
polarization state from the single image in which the zero-intensity point
of the optical vortex beam corresponds to a stereographic projection point
of the Poincar\'{e} sphere. We experimentally perform single-setup
weak measurement to validate the stereographical relationship between the
polarization state on the Poincar\'{e} sphere and the location of
the zero-intensity point.
\end{abstract}

\maketitle
\section{INTRODUCTION}
Weak measurement was originally proposed by Aharonov, Albert, and
Vaidman~\cite{PhysRevLett.60.1351} as an extension to the standard
von Neumann model of quantum measurement, inspired by
the two-state vector formalism of quantum mechanical
measurement~\cite{PhysRev.134.B1410}. This formalism is
characterized by the pre- and post-selected states of the measured system.
In contrast to an ideal (strong) measurement, weak measurement
extracts very little information about the measured system from a single
outcome, but the measured state does not collapse.
Although each outcome of weak measurement includes a
large uncertainty, the averaged value in multiple trials can build
up a significant value, called a weak value, without state collapse.
This feature makes weak measurement an ideal tool for examining
the fundamentals of quantum physics, such as quantum
correlation and quantum dynamics; for instance, see the
reviews~\cite{AR,AV,AT,Shikano,Dressel}.

On the basis of the above property, recent works have theoretically
shown that weak measurement can be used for
observing the quantum wavefunction~\cite{PhysRevA.81.012103,PhysRevA.84.052106,wu2013state,PhysRevA.88.042114,arXiv:1310.6206}.
In contrast to conventional
quantum-state tomography~\cite{QST}, this scheme records the complex-valued
weak values describing the wavefunction of the quantum state and requires
less post-processing. Recently, Lundeen and his colleagues demonstrated
weak measurement of the one-dimensional transverse wavefunction~\cite{lundeen2011direct}.
Since then, state tomography via weak measurement has been applied to
several physical systems, such as an average photon trajectory~\cite{kocsis2011observing,PhysRevLett.110.060406},
a polarization state~\cite{salvail2013full}, and an orbital angular momentum state~\cite{malik2013direct}.
All of these demonstrations, however, require the experimental
configuration to measure two noncommutative operators, such as
a position operator $\hat{X}$ and its momentum operator
$\hat{P}_x$, onto the probe state of the weak measurement. 

In this paper, we propose a stereographical-visualization scheme for
a polarization state by two-dimensional imaging of the weak value with
two commutative position operators. 
The key idea is to employ Laguerre--Gaussian
(LG) modes for a probe state in weak measurement, and it weakly interacts
with the measured polarization state, as proposed in Refs.~\cite{PhysRevLett.109.040401,PhysRevA.86.053805}.
The LG mode has a zero-intensity point (ZIP) in the center, which is generally
called an {\it optical vortex}~\cite{nye1974dislocations}, and several generation
methods have been proposed and
demonstrated~\cite{bazhenov1990laser,White1,White2,PhysRevA.45.8185,beijersbergen1994helical,PhysRevA.84.033813,Ando,kobayashi12:_helic,PhysRevLett.96.163905}.
Note that the relationship between the optical vortex beam and the weak
value has been pointed out from
different viewpoints~\cite{Dennis0,Dennis1,Dennis2,Dennis3,Magana-Loaiza}.
According to Refs.~\cite{PhysRevLett.109.040401,PhysRevA.86.053805}, 
we can extract information on the polarization state from a single
image as the two-dimensional location of the ZIP, which corresponds to the
stereographic projection point of the polarization state on the
Poincar\'{e} sphere.
We also experimentally verify the stereographical relationship between the
polarization state and the location of the ZIP. Furthermore, this is a practical demonstration
of weak measurement using an optical vortex beam as the probe state.

The remainder of this paper is organized as follows. In Sec. II, we show
the stereographical relationship between a polarization state on the
Poincar\'{e} sphere and a weak value of a polarization state and propose a
stereographical- visualization scheme for the polarization state by
using the weak measurement with an optical vortex beam. In Sec. III, we
experimentally implement our scheme by using a polarizing Sagnac
interferometer including a specialized half-wave plate, called $q$
plate. We demonstrate our proposed scheme and evaluate its accuracy from
the fidelity of the measurement results. Summary and discussion are
presented in Sec. IV.

\begin{figure}[b]
\begin{center}
\includegraphics[width=7.5cm]{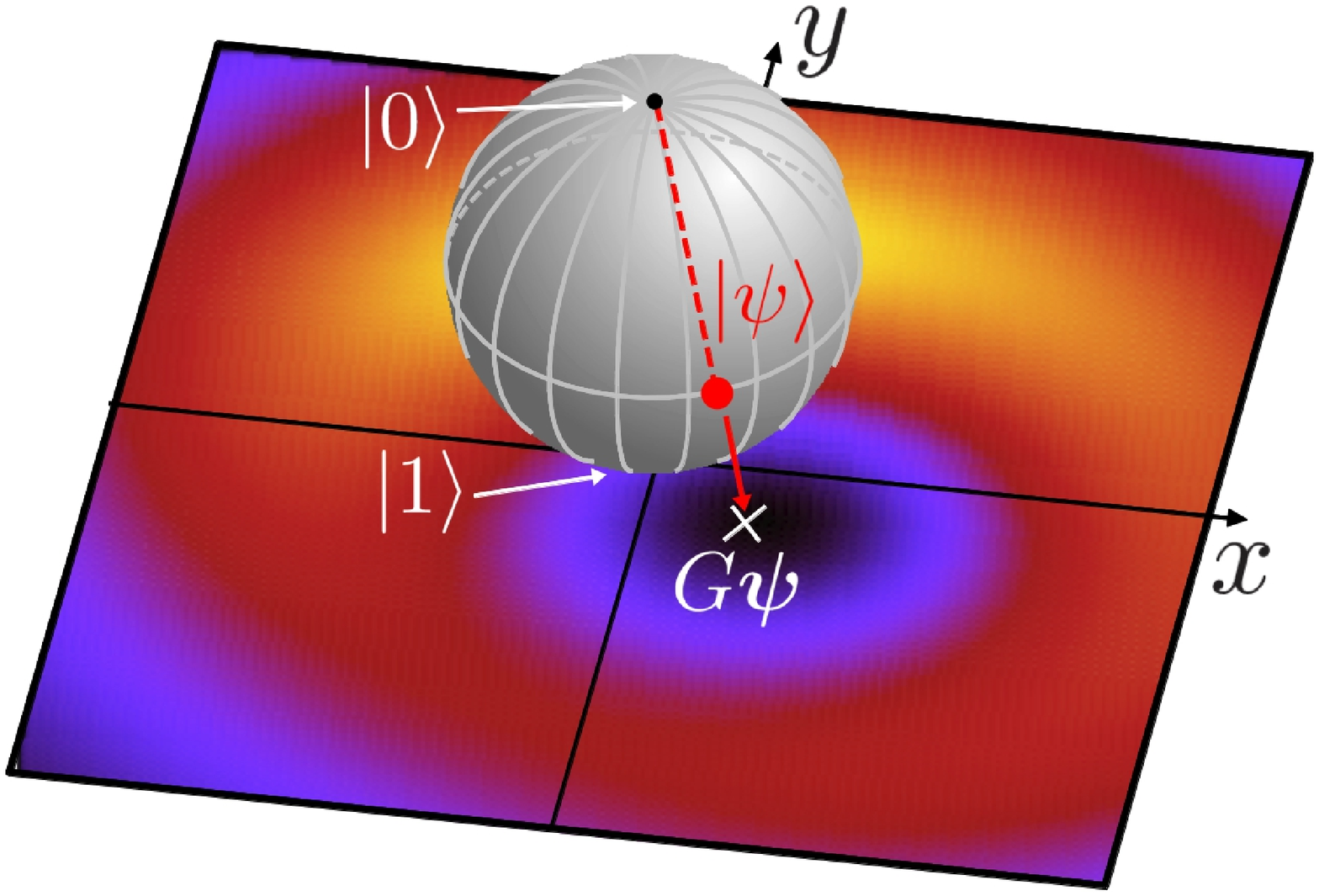}
\end{center}
\caption{Stereographic projection of qubit state $\ket{\psi}$ on Bloch sphere
 to the ZIP,
 $G \vct{\psi}= G (\Re\bracket{\hat{\sigma}_x}\sub{w},\Im\bracket{\hat{\sigma}_x}\sub{w})$.
As an example, the $l = 1$ case is illustrated.}
\label{fig:stereo_projection}
\end{figure}

\section{STEREOGRAPHICAL VISUALIZATION OF A QUBIT STATE}
Let us consider the geometrical relationship between a weak value
and a two-dimensional quantum (qubit) state. Let
any qubit state $\ket{\psi}$ spanned by orthonormal basis states $\ket{0}$ and
$\ket{1}$ be denoted as
\begin{align}
\ket{\psi}=\cos\theta\ket{0}+\ee^{\ii\phi}\sin\theta\ket{1},  \label{eq:1}
\end{align}
where $0\leq\theta\leq\frac{\pi}{2}$, and $0\leq\phi<2\pi$.
Here, the parameters $\phi$ and $\theta$ uniquely specify a point on the unit
sphere $S^2$, called the Bloch sphere, which is the same as the Poincar\'{e} sphere in optics,
in which the states $\ket{0}$ and $\ket{1}$
correspond to the north and south poles, respectively. We
consider the weak value of the observable
$\hat{\sigma}_x\equiv\ket{0}\bra{1}+\ket{1}\bra{0}$, which is the $x$ component
of the Pauli matrices, with the post-selected state $\ket{1}$:
\begin{align}
\bracket{\hat{\sigma}_x}\sub{w}
\equiv \frac{\bracketii{1}{\hat{\sigma}_x}{\psi}}
{\bracketi{1}{\psi}}
=\frac{\bracketi{0}{\psi}}
{\bracketi{1}{\psi}}
=\ee^{-\ii\phi}\cot\theta,  \label{eq:2}
\end{align}
which is generally a complex value. This weak value gives a geometrical mapping
of the qubit state on the Bloch sphere, namely, the stereographic
projection, as shown in Fig.~\ref{fig:stereo_projection}. 

The stereographic projection is one way to make a flat map of a spherical
surface. Let $P(\ket{0})$ denote the north pole on the Bloch sphere.
Given a point $P(\ket{\psi})$ related to the qubit state $\ket{\psi}$,
other than $P(\ket{0})$, the line
connecting $P(\ket{0})$ and $P(\ket{\psi})$ intersects a certain flat complex (two-dimensional real)
plane orthogonal to the line connecting $P(\ket{0})$ and $P(\ket{1})$ at
exactly one point, $G\vct{\psi} \equiv G(\Re\bracket{\hat{\sigma}_x}\sub{w},\Im\bracket{\hat{\sigma}_x}\sub{w})$.
Thus, the stereographic projection can be taken as the map
$\pi: S^2-P(\ket{0}) \rightarrow \mathbb{C} \simeq \mathbb{R}^2 : P(\ket{\psi}) \rightarrow G\vct{\psi}$.
From a geometrical viewpoint, it is easily understood that the
stereographically projected point corresponding to the weak value can become
far from the origin of the complex plane.
The south pole $P(\ket{1})$ appears at the origin of the complex plane.
Lines of latitude appear as circles around this origin.
The southern hemisphere of the Bloch sphere is not stretched very much in
the map, but the northern hemisphere is stretched quite a bit, and the
north pole is at infinity.

To extract the real and imaginary parts of the weak value from a single
image, we can use the LG mode
as the probe state~\cite{PhysRevLett.109.040401,PhysRevA.86.053805}.
The LG mode is one of the natural solutions of the paraxial wave equation
and is characterized by a radial index $p$ and an azimuthal index $l$.
The modes have an annular intensity distribution around the ZIP.
The wavefront of the LG modes is composed of $|l|$ intertwined helical
wavefronts, with a handedness given by the sign of $l$.
 In what follows, we consider the optical vortex beam with $p=0$ and
 $l>0$ for simplicity.

The amplitude distribution of the LG mode is given as
\begin{align}
\phi\sub{i}(x,y)
&=\bracketi{x,y}{\phi\sub{i}}  \label{eq:3} \\
&\propto f_l(x,y)\exp\left(-\frac{x^2+y^2}{4W_0^2}\right),  \label{eq:4}
\end{align}
where $W_0$ corresponds to the beam width,
and $f_l(x,y)\equiv (x+\ii y)^l$ determines the
ZIP as the origin. Unlike a fundamental Gaussian mode ($l = 0$),
the LG mode is no longer factorable in two directions, $x$ and $y$,
and this is a key factor for retrieving the
real and imaginary parts of the weak value from a single image. To apply the weak
measurement scheme, we use the von Neumann interaction Hamiltonian,
\begin{align}
\hat{H}=g\delta(t-t_0)\hat{\sigma}_x\otimes\hat{P}_x,  \label{eq:5}
\end{align}
where the coupling constant $g$ is sufficiently small, and $\hat{P}_x$ is the
momentum observable on the probe system conjugate to the commuting
position observable $\hat{X}$. For simplicity, we assume the interaction
to be impulsive at time $t=t_0$. After the interaction between the system
state and the LG mode probe states, we
post-select the system in $\ket{1}$, resulting
in the probe state described as
\begin{align}
\ket{\phi\sub{f}}&=\bracketii{1}
{\ee^{-\ii G\hat{\sigma}_x\otimes\hat{P}_x}}
{\psi\sub{i}}\ket{\phi\sub{i}}  \notag \\
&\simeq\bracketi{1}{\psi\sub{i}}
\exp\left(
-\ii G\bracket{\hat{\sigma}_x}\sub{w}\hat{P}_x
\right)
\ket{\phi\sub{i}}  \label{eq:6}
\end{align}
with $G\equiv g/\hbar$. Note that this approximation is justified under the condition
\begin{align}
\frac{W_0}{G}\gg\max\left(1,|\bracket{\hat{\sigma}_x}\sub{w}|\right).  \label{eq:7}
\end{align}
Without this condition, the exact calculation
can be described (see Appendix A). From Eq.~(\ref{eq:6}), the spatial intensity distribution
of the probe state after post-selection
can be calculated as
\begin{align}
|\phi\sub{f}(x,y)|^2\propto&
|\phi\sub{i}(x-G  \bracket{\hat{\sigma}_x}\sub{w},y)|^2  \notag \\
\propto&
|f_l(x-G\Re\bracket{\hat{\sigma}_x}\sub{w},y-G\Im\bracket{\hat{\sigma}_x}\sub{w})|^2\notag\\
&\hspace*{0.1cm}
\times\exp\left[-\frac{(x-G\Re\bracket{\hat{\sigma}_x}\sub{w})^2+y^2}{2W_0^2}\right].  \label{eq:8}
\end{align}
Therefore, the ZIP is shifted by the real and imaginary parts of the weak
value $\bracket{\hat{\sigma}_x}\sub{w}$ in the two-dimensional image.
Consequently, we can determine the qubit state on the Bloch sphere from the
ZIP related to the stereographic projection, as shown in Fig.~\ref{fig:stereo_projection}.

It is remarkable that our scheme can be generalized to the mixed state case, in which the qubit state
is inside the Bloch sphere. The above procedure is applied to different post-selected states,
e.g., $\ket{0}$, $(\ket{0} + i\ket{1})/\sqrt{2}$, and $(\ket{0} - i \ket{1})/\sqrt{2}$.
Then, we can always choose at least two imaging planes to satisfy
the approximation condition (\ref{eq:7}). From the two ZIPs,
we can find the crossing point on the stereographic projection, which is the qubit state in the
mixed state case. According to Refs.~\cite{PhysRevA.70.052321, PhysRevA.81.042109},
we need to prepare at least four imaging planes for any unknown qubit state.

\section{EXPERIMENT BY WEAK MEASUREMENT WITH LAGUERRE-GAUSSIAN MODES}
\begin{figure}[t]
\begin{center}
\includegraphics[width=8cm]{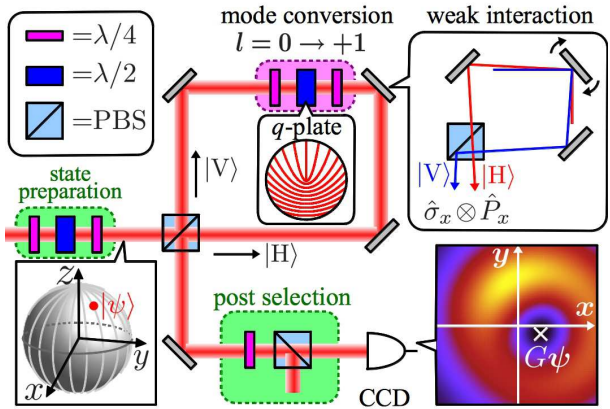}
\end{center}
\caption{Experimental setup with the polarizing
Sagnac interferometer.}
\label{fig:polarization_weak_measurement_setup}
\end{figure}
Figure~\ref{fig:polarization_weak_measurement_setup} shows our
experimental setup for demonstrating direct mapping of the polarization
state onto the ZIP.
We use a fiber-pigtailed continuous-wave laser with a wavelength of 785 nm, a power of
5 mW, and a beam waist radius of 1 mm.
First, we prepare the fundamental Gaussian beam ($l=0$) with
an arbitrary polarization state using
half-wave plates and quarter-wave plates in the state preparation stage.
Next, we convert the spatial mode to the LG mode using a
specially fabricated wave plate called a $q$ plate~\cite{PhysRevLett.96.163905}.
The $q$ plate has a suitably patterned transverse optical axis to couple
the photon spin (polarization) with its orbital angular momentum~\cite{PhysRevLett.96.163905}.
Here, we use a $q$ plate (Altechna, RPC-800-6) with a singularity charge $q=1/2$ and
uniform birefringent retardation $\delta=\pi$. This $q$ plate can modify the
mode number $l$ to $l \pm 1$, the sign of which depends on the input polarization.
The helicity of the output circular polarization is also inverted.

To implement the above mode conversion without changing the output polarization,
we propose a mode conversion setup based on a polarizing
Sagnac interferometer (PSI) including the $q$ plate.
This setup ideally is 100\% efficient for mode conversion
and has good stability because of self-compensation of the
optical paths inside the interferometer.
The incident polarization remains unchanged because of the
same optical path in the right-circulating and left-circulating paths inside the interferometer.
A single polarizing beam splitter (PBS) is used as the entry and exit gates
of the device. The PBS splits the incident beam into its horizontal component
$\ket{\text{H}} \equiv (\ket{0} + \ket{1})/\sqrt{2}$ and
its vertical component $\ket{\text{V}} \equiv (\ket{0} - \ket{1})/\sqrt{2}$,
which circulate inside the interferometer along the same path, but in
opposite directions. The spatial mode of each polarization component is converted to the LG
mode with $l=1$ by using quarter-wave plates with an angle of
$45\degree$ and the $q$ plate.
After being reflected by the mirrors, they recombine
again and exit the interferometer from the other side of the PBS.
After passing through the interferometer, the two counter-propagating
orthogonal polarizations $\ket{\text{H}}$ and $\ket{\text{V}}$ gain the same
optical length. In this way, our setup can convert the incident
fundamental Gaussian mode ($l=0$) to the LG mode ($l=1$) without a
polarization change.

Inside the PSI, the weak interaction is obtained by slightly tilting the mirror of the PSI
as expressed in Eq.~(\ref{eq:5}). After it passes through the PSI,
we post-select the polarization state into the right circular
polarization basis $\ket{\text{R}} \equiv \ket{1}$ and observe
the spatial intensity distribution using a CCD image sensor
with a pixel size of $6.45 \, \U{um}$.
Recall that our setup differs from the measurement interaction of the optical paths
in the Sagnac interferometer ~\cite{PhysRevLett.102.173601}.
Note also that this extends previous experimental studies on weak
measurement of the polarization~\cite{PhysRevLett.66.1107,pryde2005measurement}.

\begin{figure*}
\begin{minipage}[t]{9cm}
\begin{flushleft}
(a)
\end{flushleft}
\vspace*{-0.8cm}
\includegraphics[height=2cm]{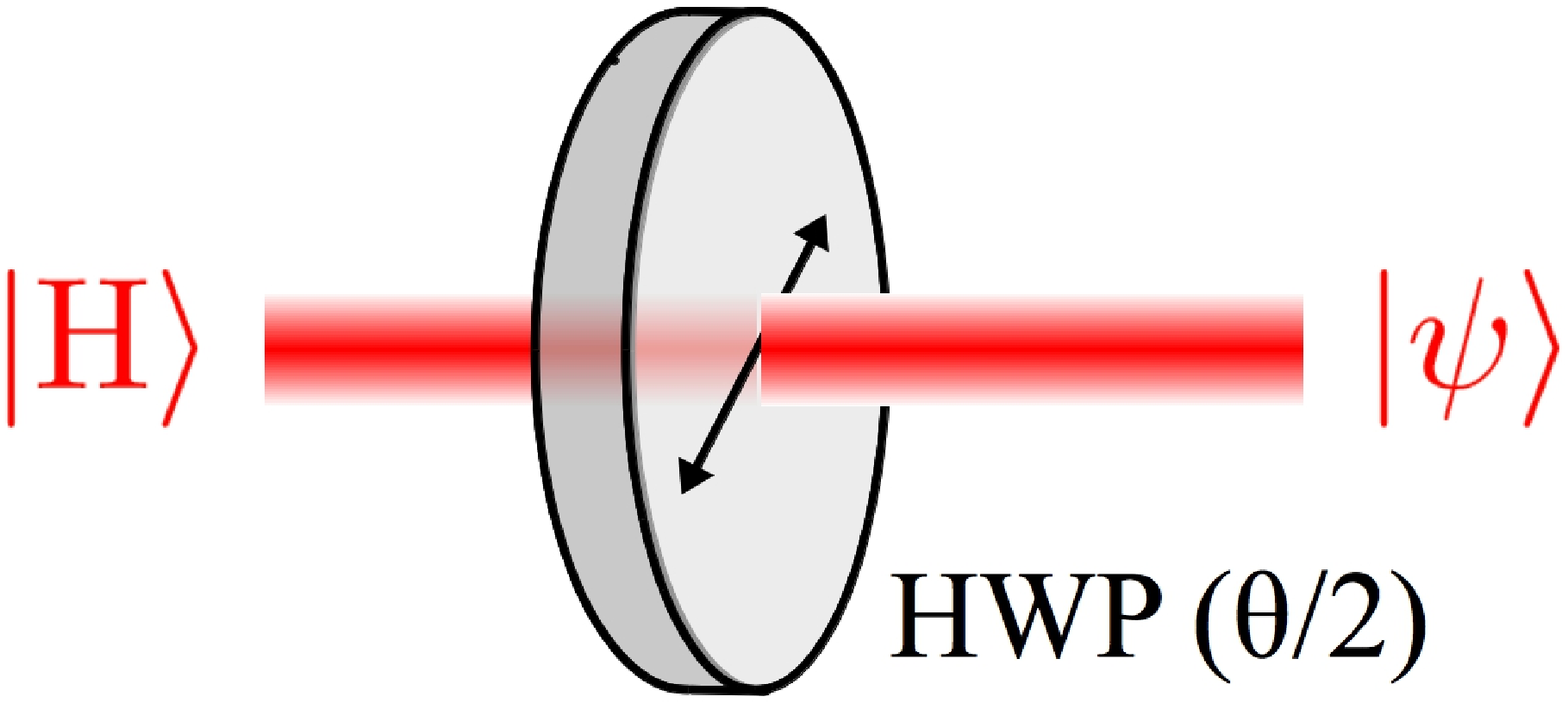}
\begin{flushleft}
\vspace*{-0.4cm}
(b)
\end{flushleft}
\vspace*{-0.2cm}
\includegraphics[width=8.5cm]{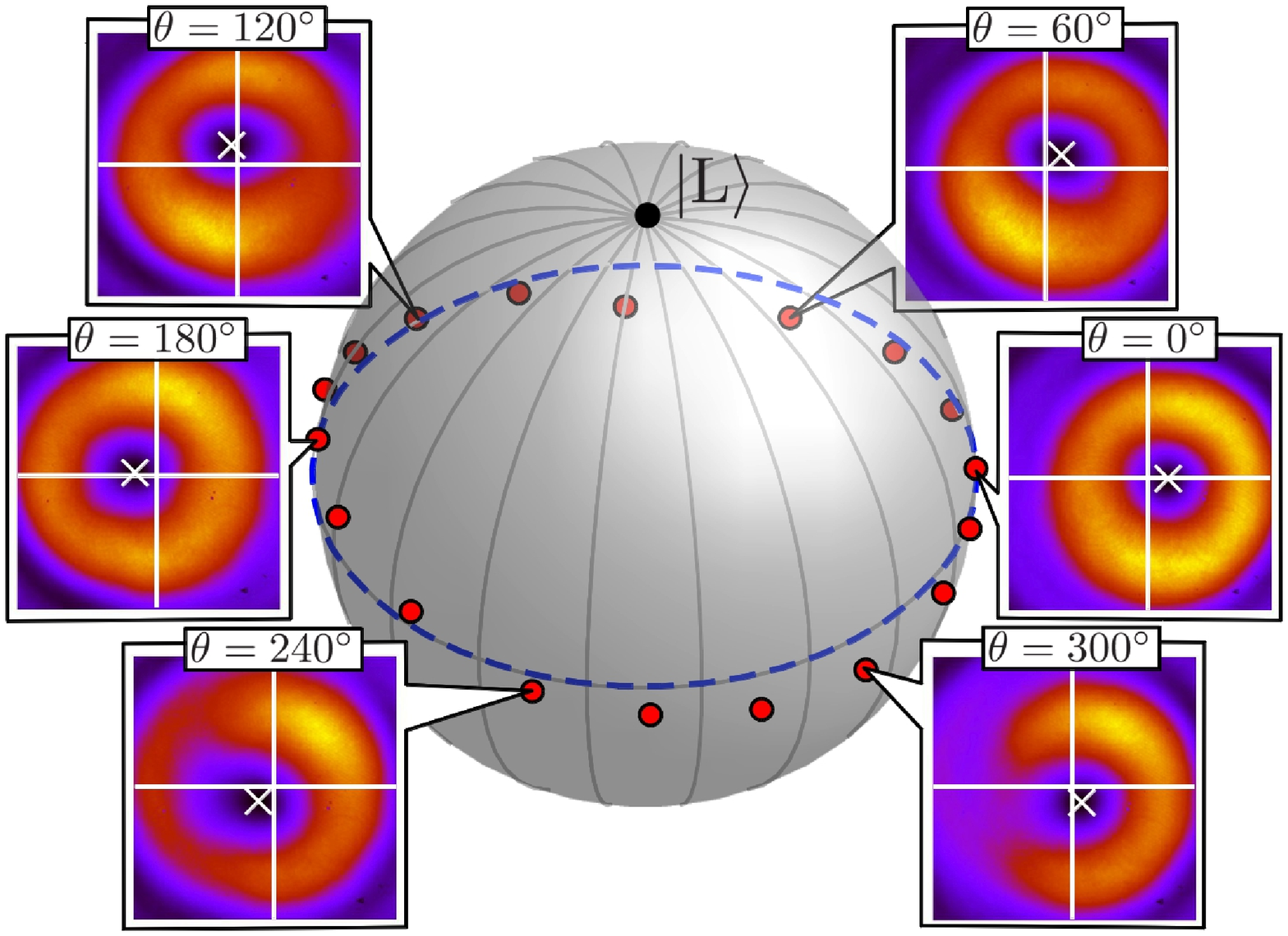}
\end{minipage}
\begin{minipage}[t]{8.5cm}
\begin{flushleft}
(c)
\end{flushleft}
\vspace*{-0.8cm}
\includegraphics[height=2cm]{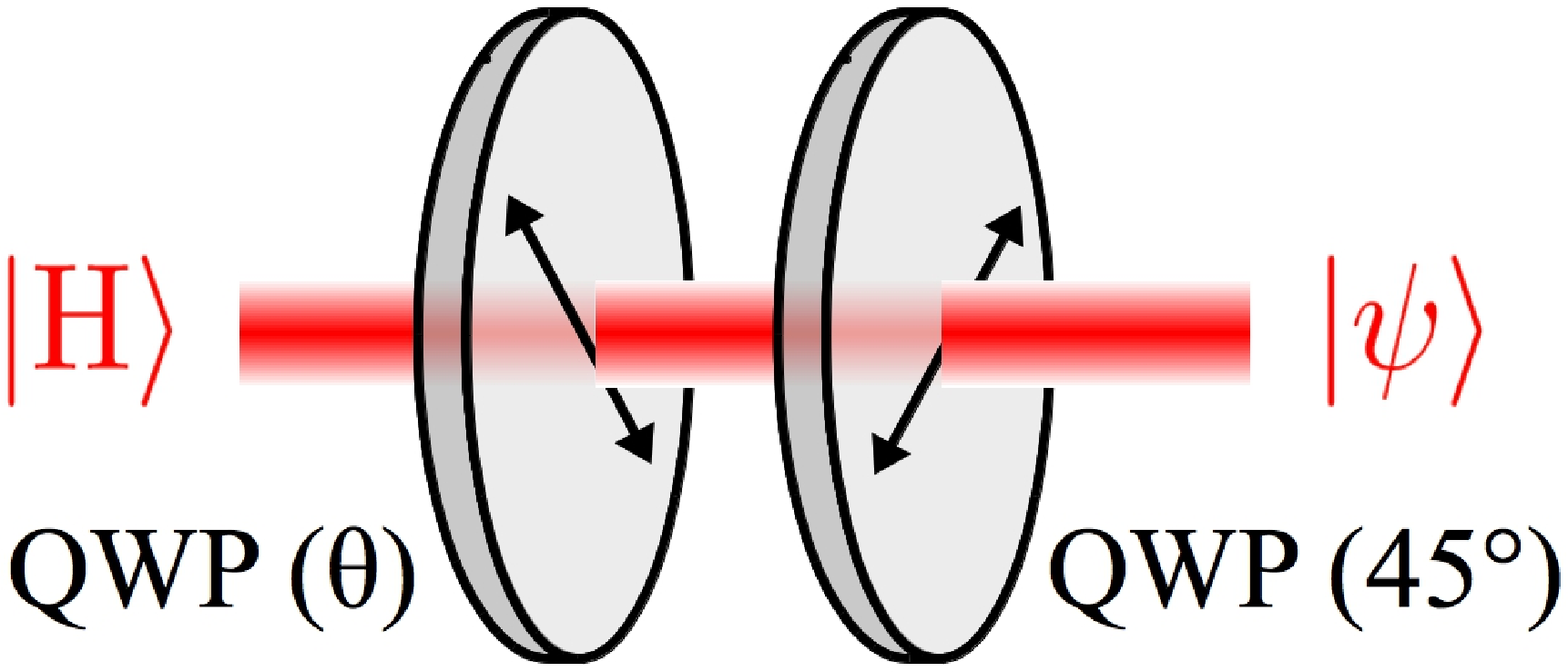}
\begin{flushleft}
\vspace*{-0.4cm}
(d)
\end{flushleft}
\vspace*{-0.2cm}
\includegraphics[width=8.5cm]{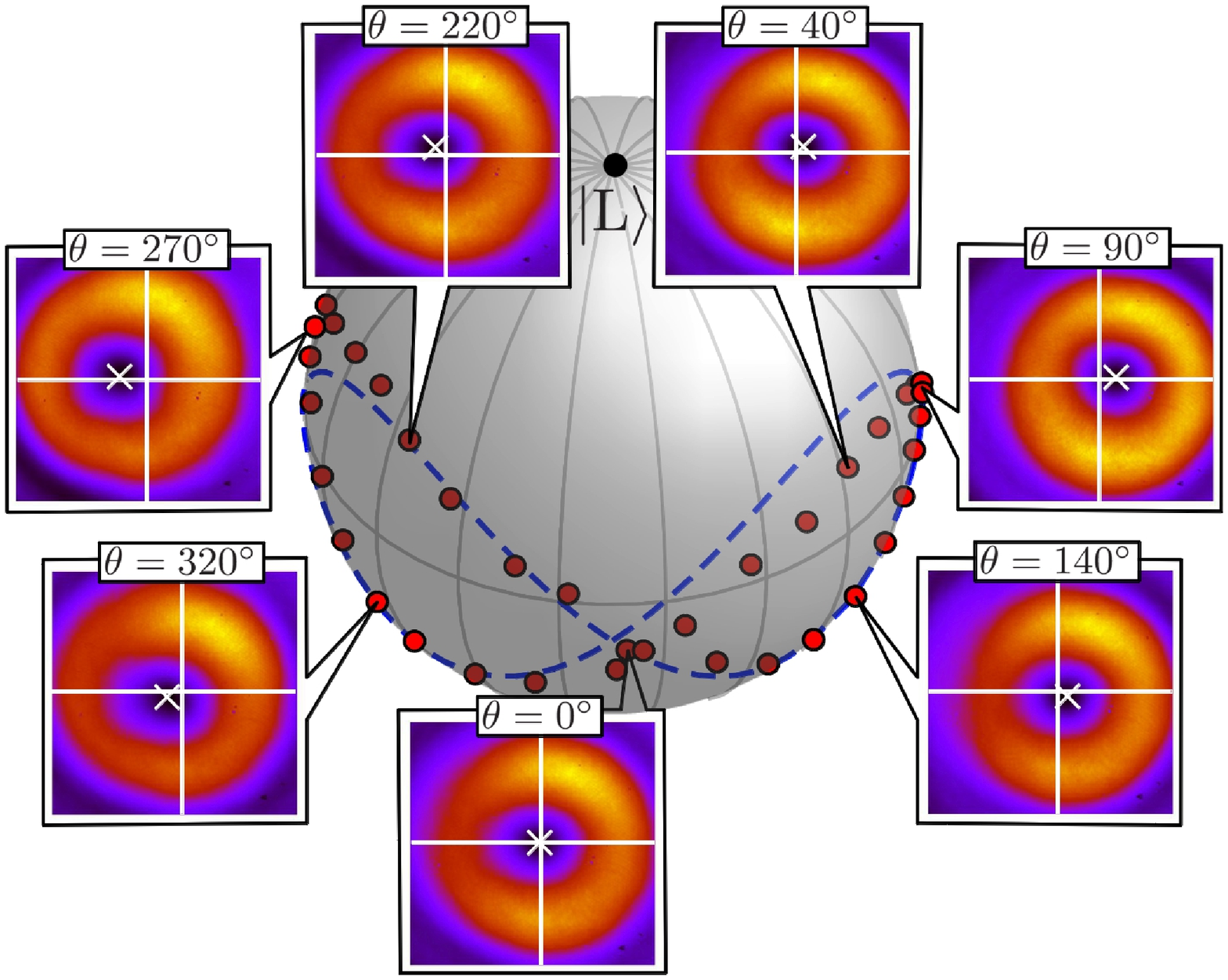}
\end{minipage}
\caption{Experimental observation of ZIP related to the stereographic projection of
 the polarization states around (a),(b) the equator and (c),(d) the
 $\infty$-shaped path of the southern hemisphere on the Poincar\'{e} sphere.
(a), (c) Experimental setup for state preparation. 
(b), (d) Theoretical path of prepared polarization statess (blue dashed
 line) and experimentally-obtained polarization states from the weak
 measurement (red points)~\cite{calibration}. Inset figures show false-color plots of
 observed intensity distribution, in which the two-dimensional location of ZIP (white
 cross mark) is determined from the weighted-averaged position within 1\% of the maximum
 intensity.}
\label{fig:results0}
\end{figure*}

\begin{figure}
\begin{center}
\includegraphics[width=8.8cm]{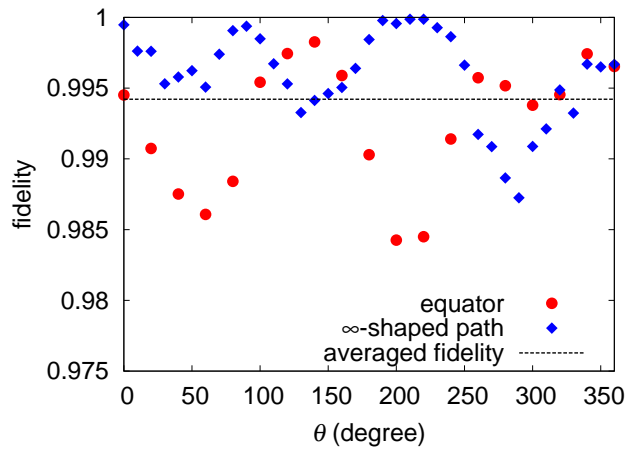}
\end{center}
\caption{Calculated fidelities of estimated states from the weak
 measurement with prepared states on the equator (red
 circles) and on the $\infty$-shaped path (blue diamonds). 
Averaged fidelity is calculated as 0.994.}
\label{fig:fidelity}
\end{figure}

Figure \ref{fig:results0} shows our experimental results.
Figures \ref{fig:results0}(a) and \ref{fig:results0}(c) show experimental setups to
prepare the polarization states. 
Figures \ref{fig:results0}(b) and \ref{fig:results0}(d) depict the observed intensity
distribution (inset figures), estimated polarization states from the weak measurement (red points), and the
theoretical path of prepared polarization states (blue dashed line).
First, we observed the linear polarization states around the equator on the
Poincar\'{e} sphere, as shown in Fig.~\ref{fig:results0}(a) and \ref{fig:results0}(b). 
We can confirm that the ZIP represented by the
white cross in Fig.~\ref{fig:results0}(b) moves along the circle corresponding to the
equator on the Poincar\'{e} sphere. 
Next, we observed the polarization states along the more complex path,
namely, the $\infty$-shaped path on the southern hemisphere of the Poincar\'{e} sphere.
The $\infty$-shaped path is realized by using two quarter-wave plates,
one of which is rotated and the other of which is fixed at an angle of
45$\degree$ with respect to the direction of the input linear
polarization [see Fig.~\ref{fig:results0}(c)]. 
We can see that the experimentally-obtained ZIP moves along
the $\infty$-shaped path, as shown in Fig.~\ref{fig:results0}(d).
These experimental results indicate that the polarization
states can be directly determined by the ZIP from a
single image of the intensity distribution.
Figure \ref{fig:fidelity} shows calculated fidelities between estimated
polarization states from the weak measurement and prepared
states determined by the angle $\theta$ [see Figs.~\ref{fig:results0}(a)
and \ref{fig:results0}(c)]. The averaged fidelity is calculated as 0.994. The measurement
error is mainly attributed to the imperfection
of the PSI setup and the mode conversion. 

\section{SUMMARY AND DISCUSSION}
In summary, we proposed a stereographical-visualization scheme for
the qubit state by two-dimensional imaging of the weak value.
We experimentally performed a single-setup weak measurement to
evaluate the polarization state of photons using the LG mode. 
We also estimated the accuracy of our scheme from the fidelity of
measurement results. 
Our approach to polarization measurement has the advantage in optics
that we may extract information on the polarization from a single image,
in cases of very small photon number, and in the characterization
of polarization-entangled quantum states. By inferring the probability
density of the polarization conditioned on \textit{a priori} information,
we can explore the implications of polarization measurements
at very low light levels. For a very small number of independent photons,
our formalism suggests that each photon represents an individual stochastic
event described by a spatial probability density function. 
However, the angle between
the ZIP and our stereographic projection
onto the complex projective plane $\mathbb{C}\mathbf{P}(1)$ of the
quantum state is related to the fidelity via the Fubini--Study
metric~\cite{RepMathPhys.9.273}. Further, the fidelity of
the mixed state case~\cite{JModOpt.41.2315} can be generalized
via the Bures metric~\cite{ChrisFuchs_PhD_dissertation}.
Therefore, the ZIP may directly visualize the
geometrical structure of the quantum state, the
K\"{a}hler manifold~\cite{geometry_of_quantum_state}.

\begin{acknowledgements}
We thank Ian Wamsley, Andrew Briggs, George C. Knee, and
Erik M. Gauger for providing useful comments and suggestions.
This work was supported by a Grant for Basic Science Research Projects
from The Sumitomo Foundation, a grant from Matsuo Foundation, IMS Joint
 Study Program, and JSPS KAKENHI
Grants No. 25790068 and No. 25287101.
\end{acknowledgements}

\begin{appendix}
\section{Exact calculation of weak measurement}
In the main text, we use the weak measurement scheme 
under the approximation condition (weak condition);
\begin{align}
\frac{W_0}{G}\gg\max\left(1,|\bracket{\hat{\sigma}_x}\sub{w}|\right).  \label{eq:9}
\end{align}
In what follows, the probe state is calculated without the weak condition 
under the same setup in the main text. 
The initial state of the probe state is set the LG beam with $l = 1$;
\begin{align}
\phi\sub{i}(x,y)
:= N ( x+i y)
\exp\left(-\frac{x^2+y^2}{4 W_0^2}\right),  \label{eq:10}
\end{align}
where $N$ is the normalization constant. 
The probe state after the measurement interaction with taking the postselection  is 
\begin{align}
\phi\sub{f}(x,y) &=\frac{\bracketi{1}{\psi}}{2}
\Bigl\{
\left(1-\bracket{\hat{\sigma}_x}\sub{w}\right)\phi\sub{i}(x+G,y)\notag\\
&\hspace*{2cm}+ \left(1+\bracket{\hat{\sigma}_x}\sub{w}\right)\phi\sub{i}(x-G,y)
\Bigr\}.  \label{eq:11}
\end{align}
The above equation can be approximated under the weak condition as
\begin{align}
\phi\sub{f}(x,y)&\simeq\bracketi{1}{\psi}
\phi\sub{i}(x-G  \bracket{\hat{\sigma}_x}\sub{w},y)
\label{eq:12}
\end{align}
Without the approximation, the average position of the intensity distribution is calculated as
\begin{align}
\bracket{\hat{X}}\sub{f}&=\frac{G\Re\bracket{\hat{\sigma}_x}\sub{w}}
{\displaystyle\frac{1}{2}\left\{1+|\bracket{\hat{\sigma}_x}\sub{w}|^2
+\eta (1-|\bracket{\hat{\sigma}_x}\sub{w}|^2)
\right\}} \notag\\
&=\frac{G\cos\phi\sin\theta}{1- \eta \cos\theta} \notag \\
&\xrightarrow{\text{weak condition}}
G\Re\bracket{\hat{\sigma}_x}\sub{w}  \label{eq:13}
\\
\bracket{\hat{Y}}\sub{f}&=\frac{G\ee^{-G^2/2W_0^2}
\Im\bracket{\hat{\sigma}_x}\sub{w}}
{\displaystyle\frac{1}{2}\left\{1+|\bracket{\hat{\sigma}_x}\sub{w}|^2
+\eta (1-|\bracket{\hat{\sigma}_x}\sub{w}|^2)
\right\}} \notag\\
&=\frac{G\ee^{-G^2/2W_0^2}
\sin\phi\sin\theta}
{1- \eta \cos\theta}  \notag \\
&\xrightarrow{\text{weak condition}}
G\Im\bracket{\hat{\sigma}_x}\sub{w},  \label{eq:14}
\end{align}
where
\begin{align}
\ket{\psi} & \equiv \cos\theta\ket{0}+\ee^{\ii\phi}\sin\theta\ket{1}, \notag \\
\eta & \equiv \left(
1-\frac{G^2}{2W_0^2}\right)\ee^{-G^2/2W_0^2}.  \label{eq:15}
\end{align}
The results under the weak condition [(\ref{eq:13}), (\ref{eq:14})] 
are the same as those in
Refs.~\cite{PhysRevLett.109.040401,PhysRevA.86.053805}.
\end{appendix}


\begin{thebibliography}{20}
\expandafter\ifx\csname natexlab\endcsname\relax\def\natexlab#1{#1}\fi
\expandafter\ifx\csname bibnamefont\endcsname\relax
  \def\bibnamefont#1{#1}\fi
\expandafter\ifx\csname bibfnamefont\endcsname\relax
  \def\bibfnamefont#1{#1}\fi
\expandafter\ifx\csname citenamefont\endcsname\relax
  \def\citenamefont#1{#1}\fi
\expandafter\ifx\csname url\endcsname\relax
  \def\url#1{\texttt{#1}}\fi
\expandafter\ifx\csname urlprefix\endcsname\relax\def\urlprefix{URL }\fi
\providecommand{\bibinfo}[2]{#2}
\providecommand{\eprint}[2][]{\url{#2}}

\bibitem{PhysRevLett.60.1351}
Y. Aharonov, D. Z. Albert, and L. Vaidman,
Phys. Rev. Lett. {\bf 60}, 1351 (1988).

\bibitem{PhysRev.134.B1410}
Y. Aharonov, P. G. Bergmann, and J. L. Lebowitz,
Phys. Rev. {\bf 134}, B1410 (1964).

\bibitem{AR}
Y. Aharonov and D. Rohrlich,
{\it Quantum Paradoxes} (Wiley-VCH, Weibheim, 2005).

\bibitem{AV}
Y. Aharonov and L. Vaidman,
in {\it Time in Quantum Mechanics}, Vol. 1,
edited by J. G. Muga, R. Sala Mayato, and I. L. Egusquiza (Springer, Berlin Heidelberg, 2008), p. 399.

\bibitem{AT}
Y. Aharonov and J. Tollaksen,
in {\it  Visions of Discovery: New Light on Physics, Cosmology, and Consciousness},
edited by R. Y. Chiao, M. L. Cohen, A. J. Leggett, W. D. Phillips, and C. L. Harper, Jr. (Cambridge University Press, Cambridge, 2011), p. 105.

\bibitem{Shikano}
Y. Shikano, in {\it Measurements in Quantum Mechanics}, edited by
        M. R. Pahlavani (InTech, Rijeka, Croatia, 2012), p. 75, arXiv:1110.5055.

\bibitem{Dressel}
J. Dressel, M. Malik, F. M. Miatto, A. N. Jordan, and R. W. Boyd,
Rev. Mod. Phys. {\bf 86}, 307 (2014).

\bibitem{PhysRevA.81.012103}
H. F. Hofmann,
Phys. Rev. A {\bf 81}, 012103 (2010).

\bibitem{PhysRevA.84.052106}
S. Massar and S. Popescu,
Phys. Rev. A {\bf 84}, 052106 (2011).

\bibitem{wu2013state}
S. Wu, Sci. Rep. {\bf 3}, 1193 (2013).

\bibitem{PhysRevA.88.042114}
A. Di Lorenzo,
Phys. Rev. A {\bf 88}, 042114 (2013).

\bibitem{arXiv:1310.6206}
L. Maccone and C. C. Rusconi, Phys. Rev. A {\bf 89}, 022122 (2014).

\bibitem{QST}
J. B. Altepeter, D. F. V. James, and P. G. Kwiat,
Lect. Notes Phys. {\bf 649}, 113 (2004).

\bibitem{lundeen2011direct}
J. S. Lundeen, B. Sutherland, A. Patel, C. Stewart, and
C. Bamber, Nature (London) {\bf 474}, 188 (2011).

\bibitem{kocsis2011observing}
S. Kocsis, B. Braverman, S. Ravets, M. J. Stevens, R. P.
Mirin, L. K. Shalm, and A. M. Steinberg, Science {\bf 332},
1170 (2011).

\bibitem{PhysRevLett.110.060406}
B. Braverman and C. Simon, Phys. Rev. Lett. {\bf 110},
060406 (2013).

\bibitem{salvail2013full}
J. Z. Salvail, M. Agnew, A. S. Johnson, E. Bolduc,
J. Leach, and R. W. Boyd, Nat. Photonics {\bf 7}, 316
(2013).

\bibitem{malik2013direct}
M. Malik, M. Mirhosseini, M. P. Lavery, J. Leach, M. J.
Padgett, and R. W. Boyd, Nat. Commun. {\bf 5}, 3115 (2014).

\bibitem{PhysRevLett.109.040401}
G. Puentes, N. Hermosa, and J. P. Torres, Phys. Rev.
Lett. {\bf 109}, 040401 (2012).

\bibitem{PhysRevA.86.053805}
H. Kobayashi, G. Puentes, and Y. Shikano, Phys. Rev.
A {\bf 86}, 053805 (2012).

\bibitem{nye1974dislocations}
J. Nye and M. Berry, Proc. R. Soc.
London, Ser. A {\bf 336}, 165 (1974).

\bibitem{bazhenov1990laser}
V. Y. Bazhenov, M. Vasnetsov, and M. Soskin, JETP Lett.
{\bf 52}, 429 (1990).

\bibitem{White1}
A. G. White, C. P. Smith, N. R. Heckenberg, H. Rubinsztein-Dunlop, R. McDuff,
C. O. Weiss, and C. Tamm,
J. Mod. Opt. {\bf 38}, 2531 (1991).

\bibitem{White2}
N. R. Heckenberg, R. McDuff, C. P. Smith, and A. G. White,
Opt. Lett. {\bf 17}, 221 (1992).

\bibitem{PhysRevA.45.8185}
L. Allen, M. W. Beijersbergen, R. J. C. Spreeuw, and J. P. Woerdman,
Phys. Rev. A
{\bf 45}, 8185 (1992).

\bibitem{beijersbergen1994helical}
M. Beijersbergen, R. Coerwinkel, M. Kristensen, and
J. Woerdman, Opt. Commun. {\bf 112}, 321 (1994).

\bibitem{PhysRevLett.96.163905}
L. Marrucci, C. Manzo, and D. Paparo, Phys. Rev. Lett.
{\bf 96}, 163905 (2006).

\bibitem{Ando}
T. Ando, Y. Ohtake, N. Matsumoto, T. Inoue, and N. Fukuchi,
Opt. Lett. {\bf 34}, 34 (2009).

\bibitem{PhysRevA.84.033813}
M. Mansuripur, A. R. Zakharian, and E. M. Wright,
Phys. Rev. A {\bf 84}, 033813 (2011).

\bibitem{kobayashi12:_helic}
H. Kobayashi, K. Nonaka, and M. Kitano, Opt. Express
{\bf 20}, 14064 (2012).

\bibitem{Dennis0}
M. R. Dennis and J. B. G\"{o}tte,
New J. Phys. {\bf 14}, 073013 (2012).

\bibitem{Dennis1}
J. B. G\"{o}tte and M. R. Dennis,
New J. Phys. {\bf 14}, 073016 (2012).

\bibitem{Dennis2}
M. R. Dennis and J. B. G\"{o}tte,
Phys. Rev. Lett. {\bf 109}, 183903 (2012).

\bibitem{Dennis3}
J. B. G\"{o}tte and M. R. Dennis,
Opt. Lett. {\bf 38}, 2295 (2013).

\bibitem{Magana-Loaiza}
O. S. Maga\~{n}a-Loaiza, M. Mirhosseini, B. Rodenburg, and R. W. Boyd, 
arXiv:1312.2981.

\bibitem{PhysRevA.70.052321}
J. \v{R}eh\'{a}\v{c}ek, B.-G. Englert, and D. Kaszlikowski,
Phys. Rev. A {\bf 70}, 052321 (2004).

\bibitem{PhysRevA.81.042109}
J. Nunn, B. J. Smith, G. Puentes, I. A. Walmsley, and J. S. Lundeen
Phys. Rev. A {\bf 81}, 042109 (2010).

\bibitem{PhysRevLett.102.173601}
P. B. Dixon, D. J. Starling, A. N. Jordan, and J. C. Howell,
Phys. Rev. Lett. {\bf 102}, 173601 (2009).

\bibitem{PhysRevLett.66.1107}
N. W. M. Ritchie, J. G. Story, and R. G. Hulet, Phys.
Rev. Lett. {\bf 66}, 1107 (1991).

\bibitem{pryde2005measurement}
G. J. Pryde, J. L. O'Brien, A. G. White, T. C. Ralph, and H. M.
Wiseman, Phys. Rev. Lett. {\bf 94}, 220405 (2005).

\bibitem{RepMathPhys.9.273}
A. Uhlmann, Rep. Math. Phys. {\bf 9}, 273 (1976).

\bibitem{JModOpt.41.2315}
R. Jozsa, J. Mod. Opt. {\bf 41}, 2315 (1994).

\bibitem{ChrisFuchs_PhD_dissertation}
C. A. Fuchs, Ph.D. thesis, University of New Mexico, 1995, arXiv:quant-ph/9601020.

\bibitem{geometry_of_quantum_state}
I. Bengtsson and K. {\.Z}yczkowski,
{\it Geometry of Quantum States} (Cambridge University Press, Cambridge,
        2006).

\bibitem{calibration}
The weak interaction inside the PSI induces $240\U{um}$-deflection on
        the observation plane between the horizontally-polarized and
        vertically-polarized components. From this deflection value, the
        two-dimensional location of the ZIP is calibrated to the size of the Poincar\'{e} sphere. 
\end{thebibliography}
\end{document}